\documentstyle[twocolumn,seceq]{jpsj}

\def\runtitle{Hydrodynamic Equations 
in Quantum Hall Systems at Large Currents}
\def\runauthor{Hiroshi {\sc Akera}}

\title{Hydrodynamic Equations 
in Quantum Hall Systems at Large Currents\footnote{to be published  
in J.~Phys.~Soc.~Jpn. {\bf 71} (2002) No.1.}}

\author{Hiroshi {\sc Akera}\footnote{E-mail: akera@eng.hokudai.ac.jp}}

\inst{Department of Applied Physics, 
Hokkaido University, Sapporo 060-8628}

\recdate{August 31, 2001}

\abst
{Hydrodynamic equations (HDEQs) are derived which 
describe spatio-temporal evolutions of 
the electron temperature and the chemical potential  
of two-dimensional systems in strong magnetic fields 
in states with large diagonal resistivity appearing 
at the breakdown of the quantum Hall effect. 
The derivation is based on microscopic electronic processes 
consisting of drift motions in a slowly-fluctuating potential 
and scattering processes due to 
electron-electron and electron-phonon interactions. 
In contrast with the usual HDEQs, 
one of the derived HDEQs has a term with 
an energy flux perpendicular to the electric field 
due to the drift motions in the magnetic field. 
As an illustration, 
the current distribution is calculated using the derived HDEQs. 
}

\kword{integer quantum Hall effect, nonlinear transport, theory}

\begin{document}
\sloppy
\maketitle

\def\rmd{{\rm d}}
\def\vj{\mib j}
\def\tj{\tilde{j}}
\def\cj{\check{j}}
\def\tvj{\tilde{\vj}}
\def\cvj{\check{\vj}}
\def\jc1{j_{\rm c1}}
\def\vB{\mib B}
\def\vE{\mib E}
\def\tE{\tilde{E}}
\def\tEc1{\tE_{\rm c1}}
\def\Ec1{E_{\rm c1}}
\def\cE{\check{E}}
\def\tvE{\tilde{\vE}}
\def\cvE{\check{\vE}}
\def\vr{\mib r}
\def\tvr{\tilde{\vr}}
\def\vn{\mib n}
\def\vez{{\mib e}_z}
\def\vq{\mib q}
\def\td{\tilde d}
\def\vd{\mib d}
\def\tvd{\tilde {\vd}}
\def\tnabla{\tilde {\nabla}}
\def\vnabla{\mib \nabla}
\def\tvnabla{\tilde{\vnabla}}
\def\cvnabla{\check{\vnabla}}
\def\cnablay{{\partial \over \partial \check y}}
\def\ts{\tilde {\sigma}}
\def\sxx{\sigma_{xx}}
\def\sxy{\sigma_{xy}}
\def\syx{\sigma_{yx}}
\def\syy{\sigma_{yy}}
\def\tsxx{\tilde \sigma_{xx}}
\def\tsxy{\tilde \sigma_{xy}}
\def\tsyx{\tilde \sigma_{yx}}
\def\tsyy{\tilde \sigma_{yy}}
\def\tsxxs{\tilde \sigma_{xx}^{\rm s}}
\def\rxx{\rho_{xx}}
\def\tr{\tilde \rho}
\def\trxx{\tilde \rho_{xx}}
\def\tuxy{\tilde u_{xy}}
\def\tk{\tilde {\kappa}}
\def\tCe{\tilde C_{\rm e}}
\def\kB{k_{\rm B}}
\def\Te{T_{\rm e}}
\def\tTe{\tilde T_{\rm e}}
\def\tT{\tilde T}
\def\Tc1{T_{\rm c1}}
\def\tTc1{\tilde T_{\rm c1}}
\def\tmu{\tilde \mu}
\def\mur{\mu_{\rm r}}
\def\tmur{\tilde \mu_{\rm r}}
\def\tell{\tilde \ell}
\def\lvh{\ell_{\rm vh}}
\def\tlvh{\tell_{\rm vh}}
\def\lc1{\ell_{\rm c1}}
\def\tvh{\tau_{\rm vh}}
\def\tc1{\tau_{\rm c1}}
\def\tS{\tilde S}
\def\oc{\omega_{\rm c}}
\def\ve{\varepsilon}
\def\tve{\tilde{\varepsilon}}
\def\vera{\ve_{{\rm r}\alpha}}
\def\verb{\ve_{{\rm r}\beta}}
\def\tvera{\tve_{{\rm r}\alpha}}
\def\tverb{\tve_{{\rm r}\beta}}
\def\em{\ve_{\rm m}}
\def\Ch{C_{\rm h}}
\def\Cp{C_{\rm p}}
\def\eaia{\ve_{\alpha i a}}
\def\eaib{\ve_{\alpha i b}}
\def\ebkc{\ve_{\beta k c}}
\def\ebld{\ve_{\beta l d}}
\def\ebkd{\ve_{\beta k d}}
\def\teaia{\tve_{\alpha i a}}
\def\teaib{\tve_{\alpha i b}}
\def\tebkc{\tve_{\beta k c}}
\def\tebld{\tve_{\beta l d}}
\def\tebkd{\tve_{\beta k d}}
\def\faia{f_{\alpha i a}}
\def\faib{f_{\alpha i b}}
\def\fbkc{f_{\beta k c}}
\def\fbld{f_{\beta l d}}
\def\fbkd{f_{\beta k d}}
\def\tW{\tilde W}
\def\tJ{\tilde J}
\def\Wiaib
{\tW^{\rm p}_{\alpha i a\rightarrow i b}\ }
\def\Wiaibkcld
{\tW^{\alpha i a \rightarrow i b}_{\beta k c\rightarrow l d}\ }
\def\Wiaibkckd
{\tW^{\alpha i a \rightarrow i b}_{\beta k c\rightarrow k d}\ }
\def\Wiaibkcid
{\tW^{\alpha i a \rightarrow i b}_{\beta k c\rightarrow i d}\ }
\def\Jiaibkcld
{\ \tJ^{\alpha i a \rightarrow i b}_{\beta k c\rightarrow l d}\ }
\def\Jibialdkc
{\ \tJ^{\alpha i b \rightarrow i a}_{\beta l d \rightarrow k c}\ }
\def\Jiaibkckd
{\ \tJ^{\alpha i a \rightarrow i b}_{\beta k c\rightarrow k d}\ }
\def\Jibiakdkc
{\ \tJ^{\alpha i b \rightarrow i a}_{\beta k d\rightarrow k c}\ }
%++++++++++++++++++++++++++++++++++++++++++++++++++++
\section{Introduction}

The diagonal resistivity $\rxx$ is extremely small 
in the quantum Hall effect~\cite{Klitzing80,Kawaji81} (QHE) 
which occurs when the current density $j$ is small.
When $j$ is increased up to a critical value, on the other hand, 
$\rxx$ increases by several orders of magnitude 
within a narrow range of $j$, 
and the QHE breaks down~\cite{Ebert83,Cage83,Kuchar84} 
(reviews are given in refs.~\citen{Kawaji94,Nachtwei99,Komiyama00}).  
Accompanying the breakdown of the QHE, 
spatial~\cite{Komiyama96,Kawaguchi97,Kaya98,Kaya99,Kawano00} 
and temporal~\cite{Cage83,Ahlers93,Sagol01} evolutions of $\rxx$  
have been observed, 
and other various phenomena are expected 
such as pattern formations and oscillatory instabilities 
which have been extensively studied 
in bulk semiconductors.~\cite{Scholl01}  
Unique phenomena are also expected in the present systems 
with strong magnetic fields,  
which have been partly revealed in the 
experiments.~\cite{Komiyama96,Kawaguchi97,Kaya98,Kaya99,Kawano00}  

Theoretically, on the other hand, 
a hot-electron model~\cite{Ebert83,Gurevich84,Komiyama85}  
has been proposed as a mechanism of the breakdown of the QHE.
It is proposed in this model that 
the electron heating is responsible for the increase of $\rxx$ 
at the breakdown, that is,   
the electron temperature $\Te$ is the key variable 
in determining $\rxx$. 
Later the hot-electron model has been supported by 
observed spatial variations of $\rxx$ 
at a constant $j$,~\cite{Komiyama96,Kawaguchi97,Kaya98,Kaya99} 
which are understood if we assume that 
spatial variations of $\Te$ are responsible for those of $\rxx$. 
The importance of $\Te$ 
has already been suggested by the previous experiments showing 
a bistability of a small-$\rxx$ state and a large-$\rxx$ state 
at the breakdown,~\cite{Ebert83,Cage83,Kuchar84} 
since the presence of such bistability requires 
another variable, in addition to $j$, to determine $\rxx$.  
Based on the hot-electron model, 
a hydrodynamic equation (HDEQ) 
to calculate spatio-temporal variations of $\Te$
has been proposed for one-dimensional variations 
in the previous work.~\cite{Gurevich84} 
However, its validity in the present quantum Hall systems is not 
known, because it was not derived 
on the basis of microscopic electronic processes. 
The theoretical framework to predict a variety of phenomena 
in the present systems has not been established.

In this study, we derive a set of HDEQs 
to describe spatio-temporal variations of 
the electron temperature $\Te$ and the chemical potential $\mu$ 
in large-$\rxx$ states 
on the basis of microscopic processes in 
a two-dimensional electron system (2DES)  in strong magnetic fields 
in the presence of a slowly-fluctuating potential due to ionized donors. 
As an example of its applications, 
we calculate the current distribution. 
One of the HDEQs in this paper is a generalization of 
the HDEQ given in our previous paper,~\cite{Akera01}  
which is the equation for $\Te$ in the case of 
one-dimensional spatial modulations restricted to the current direction.

The organization of this paper is as follows. 
In \S 2, three major assumptions employed in deriving the HDEQs 
are introduced with several minor assumptions, and 
a model for electronic states and processes,  
which has been proposed in our previous paper,~\cite{Akera00} 
is described. 
In \S 3, 
two HDEQs are derived which 
describe evolutions of the electron density and the energy density, 
respectively. 
In \S 4, 
formulas of the electron and energy flux densities 
as well as the energy loss, 
appearing in the HDEQs, are derived. 
In \S 5, 
values of coefficients in the HDEQs are estimated, 
and scaled HDEQs are introduced. 
The critical electric field is calculated 
using the estimated values of coefficients 
in steady and uniform states at even-integer filling factors, 
and is compared with observed values.   
In \S 6, 
the HDEQs are applied to 
obtain the current distribution at even-integer filling factors. 
In \S 7, 
conclusions and discussions are given.

%++++++++++++++++++++++++++++++++++++++++++++++++++++
\section{Assumptions and Model}

\subsection{Single-Electron Approximation and Scattering Processes}

The potential $V_{\rm donor}$ produced by ionized donors fluctuates 
in the plane of the 2DES. 
Its typical length scale~\cite{Nixon90}, or 
the average distance between a potential hill and a neighboring 
valley, $\lvh$, is 
$\lvh \sim 0.1\ {\rm \mu m} \gg \ell$ with $\ell$ the magnetic length. 
$V_{\rm donor}$ is partly screened  
by the electron-electron interaction $V_{\rm ee}$. 
The screened potential calculated 
in the Hartree approximation~\cite{Wulf88,Efros88} 
has an amplitude 
approximately equal to the Landau level separation, $\hbar \oc$. 
In such a potential, 
electrons (holes) populate in the $N+1$th ($N$th) Landau level, 
even in a filling factor of $2N$ ($N=1,2,\cdots$). 

In deriving the HDEQs, 
we take into account processes which are not included 
in the Hartree approximation. 
We divide the electron-electron interaction into two terms:
$V_{\rm ee} = \left<V_{\rm ee} \right> 
+ \left( V_{\rm ee} -  \left<V_{\rm ee} \right>  \right) $. 
The first term is the electron-electron interaction averaged 
with respect to the distribution of other electrons 
in the Hartree approximation. 
$\left<V_{\rm ee} \right>$ and $V_{\rm donor}$ give 
the screened potential as calculated 
by the previous authors.~\cite{Wulf88,Efros88} 
We introduce single-electron states in such a mean-field potential.  
The second term, on the other hand, gives scattering processes between 
the single-electron states. 
We consider such scattering processes only in the lowest order 
of $V_{\rm ee} -  \left<V_{\rm ee} \right>$. 

\def\noteCB{{\bf [[ Note:}
The Coulomb blockade is neglected in transitions.  
It is reasonable since 
the charging energy is much smaller than $\pi \kB \Te$:  
$e^2/\epsilon \lvh =1$ meV and $\pi \kB \Te=10$ meV ($\Te=30$K). 
{\bf ]]}}%\noteCB

\subsubsection{Model of electronic states and processes}

Electronic states in the slowly-varying screened potential 
consist of localized states with size $\sim \lvh$
around a potential hill or valley
and extended states with energy at the center of the broadened Landau level. 
Each state has a level broadening due to the electron-electron scatterings. 
An estimate of the level broadening is 
$\Gamma\sim 0.2 \hbar\oc$ and 
that of the inelastic scattering length is $\ell_{\rm ee} \sim \ell$, 
according to the estimation in \S \ref{sec:inelastic} at $\Te \sim \Tc1$ 
(the definition of $\Tc1$ is given in \S \ref{sec:scales}).  
Due to such electron-electron scatterings 
quantized energy levels of the localized states are smeared out. 
We therefore label the localized states 
by a continuous momentum $a$ 
which is related to a continuous energy 
in terms of the local electric field. 
The same electron-electron scatterings destroy 
the phase coherence of extended states. 
In such situation it is convenient to introduce wave packets 
with size $\sim \ell_{\rm ee}$
by superposing extended states. 
Here we divide the 2DES into regions so that 
each region contains one valley or one hill 
separated by equipotential lines of the fluctuating potential 
at the energy of extended states. 
We include each wave packet 
traveling along the boundary equipotential line 
in either the neighboring hill or valley region. 
All states are then labeled by $\alpha$, $i$, and $a$ 
where $\alpha=N \sigma$ labels a Landau level 
with $N=1,2,\cdots$ the Landau index and $\sigma$ the spin, 
$i$ labels a region, and $a$ labels a momentum. 

Processes which transfer electrons between states 
consist of drift motions and scattering processes. 
Drift motions perpendicular to the local electric field 
transfer electrons in extended states 
between neighboring regions 
(which are called d-processes), 
giving the Hall current perpendicular to the macroscopic electric field.
Scattering processes we consider are due to 
electron-electron and electron-phonon interactions. 
In our model of scattering processes, 
a transition of an electron is restricted to be 
within the same level (diagonal in $\alpha$) and 
no inter-Landau-level transitions are considered, 
since the transition rates are much smaller than 
those of intra-Landau-level transitions.\cite{Akera00}  
In electron-phonon scatterings (which are called p-processes) 
we consider scatterings within the same region only (diagonal in $i$). 
In electron-electron scatterings, we consider the following two types.  
In the first type, one electron hops between neighboring regions, 
that is, between a hill and a valley, 
and another makes a transition within a region (h-processes). 
In the second type, 
both electrons make a transition within a region (s-processes).  
The temporal evolution of 
the occupation probability $\faia(t)$ of state $\alpha i a$ is given by 
the Boltzmann-type equation:  
%====================================================
\begin{eqnarray}
{\partial \faia \over  \partial t}&=& 
\left({\partial \faia \over \partial t}\right)_{\rm d}
+\left({\partial \faia \over \partial t}\right)_{\rm p} 
\nonumber \\
&+&\left({\partial \faia \over \partial t}\right)_{\rm h}
+\left({\partial \faia \over \partial t}\right)_{\rm s}
\ .
\label{eq:ft}
\end{eqnarray}
%====================================================

\subsubsection{Characteristic scales}
\label{sec:scales}

The lower critical current density $\jc1$ 
and the lower critical electric field $\Ec1$ are 
defined by the values at which the large-$\rxx$ state becomes unstable. 
Values of the current density and the electric field 
are of the order of $\jc1$ and $\Ec1$, respectively,  
in this paper. 
Similarly, values of $\Te$ are around $\Tc1$, which is the value 
at $\jc1$ in the large-$\rxx$ state. 

The length scales in the present system are
the magnetic length $\ell$, 
the average distance between a valley and a hill $\lvh$, 
and $\lc1$ defined by $e \Ec1 \lc1 =\hbar\oc$. 
$\lc1$ is shown in \S \ref{sec:scaledHDEQs} to be the length scale of 
variations of $\Te$ and $\mu$. 
The typical sizes are 
$\ell=0.01\mu{\rm m}$, $\lvh \sim 0.1\mu{\rm m}$, and 
$\lc1 \sim 2\mu{\rm m}$
at $B=5$T and $\Ec1=50$V/cm, 
and $\ell \ll \lvh \ll \lc1$. 
The corresponding time scales are 
$\omega_c^{-1}$, $\tvh=\lvh/v_{\rm vh}$, and 
$\tc1=\lc1/v_{\rm c1}$ 
where  
$v_{\rm vh}=eE_{\rm vh} \ell^2/\hbar $ 
is the group velocity 
at the local electric field between a valley and a hill 
$E_{\rm vh}=\hbar\oc/e\lvh$, 
and 
$v_{\rm c1}=e \Ec1 \ell^2/\hbar $. 
The ratios between the time scales are 
$\omega_c \tvh=\lvh^2 /\ell^2 \sim 10^2$
and $\omega_c \tc1 =\lc1^2/\ell^2 \sim 10^4$. 
Finally the energy scale is $\hbar\oc$.

\subsection{Local Equilibrium}

We assume the local equilibrium in the scale of $\lvh$ 
and introduce the electron temperature $\Te(\vr_i,t)$ and 
the chemical potential $\mu(\vr_i,t)$ 
of the $i$th region at the position $\vr_i=(x_i, y_i)$. 
In the local equilibrium $\faia(t)$ becomes 
%====================================================
\begin{equation}
\faia(t)= f(\eaia-\mu(\vr_i,t), \Te(\vr_i,t)) ,
\end{equation}
%====================================================
where 
$f(\ve, T) =1/[\exp(\ve/\kB T) +1]$
and $\eaia$ is the energy of state $\alpha i a$. 
We neglect deviations of $\faia(t)$ 
due to the presence of electron and energy fluxes. 
With use of such averaged occupation probability, 
temporal evolutions of the system become irreversible.  

\def\noteTimerev{{\bf [[ Note:}
The Boltzmann-type equation has the time reversal invariance, 
unless the averaged distribution function is employed in the collision terms. 
The HDEQ, on the other hand, 
has no time reversal invariance. 
{\bf ]]}}%\noteTimerev

To be consistent with the local equilibrium in the scale of $\lvh$, 
it is necessary that 
the inelastic scattering length $\ell_{\rm ee}$ of extended states 
should be shorter than $\lvh$. 
The estimation in \S \ref{sec:inelastic} gives 
$\ell_{\rm ee} \ll \lvh$ when $\Te \sim \Tc1$. 
Note that $\ell_{\rm ee}$ becomes much longer than $\lvh$ 
in the very-low-$\Te$ region, 
in which the local equilibrium is violated. 

\def\noteWP{{\bf [[ Note:}
We can make a wave packet by superposing states 
within $\Delta \ve =\hbar /\tau_{\rm ee}=0.2 \hbar\oc$.  
The length of the wave packet $\ell_{\rm wp}$ is given by 
$\ell_{\rm wp}\sim{ 1\over \Delta k}
={\hbar  \over \Delta \ve} {\Delta \ve \over \hbar\Delta k}
=\tau_{\rm ee} v_{\rm vh}
=\ell_{\rm ee}  \sim \ell $. 
{\bf ]]}}%\noteWP

It is also necessary to have efficient electron and energy transfers 
among different levels with different $N$ and $\sigma$ 
in order to keep the local equilibrium. 
Therefore we must restrict our discussion to the case where 
temporal variations of macroscopic variables are slow enough 
to keep the local equilibrium.

\subsection{Spatial Averaging}

The 2DES actually has a random configuration 
of potential hills and valleys 
instead of a periodic array. 
Due to such randomness, 
electron and energy densities, 
as well as electron and energy flux densities, fluctuate 
from place to place. 
We here take an average of these quantities  
over the area larger than the scale of fluctuating potential, $\lvh$. 
Owing to this averaging, 
the equations become translationally invariant. 
To justify this averaging,  
it is necessary that $\lvh$ is much shorter than 
the length scale $\ell_{\rm var}$ of variations of $\Te$ and $\mu$. 
This is approximately satisfied, 
since it is shown in \S \ref{sec:scaledHDEQs} 
that $\ell_{\rm var} \sim \lc1$. 
Using $\lvh \ll \ell_{\rm var}$, 
the difference of a quantity between neighboring regions 
is replaced with the differential in \S \ref{sec:terms}.   

The averages of the quantities depend in general on 
the distribution function of the distance between a hill and a valley, 
which is mainly determined by the average $\lvh$
and the standard deviation. 
We neglect effects of the standard deviation 
in the calculation in the present paper.

\subsection{Other Assumptions}

We also make minor assumptions in order to simplify the HDEQs,  
although it may not be difficult to go beyond them. 

(a) We neglect the change of electronic states 
(in the single-electron approximation) 
with the applied electric field 
since $\Ec1 \sim 0.1 E_{\rm vh}$.  

(b) The screened potential depends on $\Te$ 
because the local electron density depends on $\Te$. 
We neglect such dependence
by considering only the temperature range around $\Tc1$. 

(c) We neglect the broadening of energy spectra of extended states 
due to electron-electron scattering processes 
and the resulting decrease of the activation energy,  
which are of higher order in electron-electron interactions. 

(d) In addition to the valley-hill hopping processes 
which we have introduced above,
elastic tunneling processes giving valley-valley and hill-hill 
transitions of an electron also contribute to 
the current along the electric field and consequently to $\rxx$. 
We neglect such tunneling processes in this paper, for simplicity. 

(e) In calculating terms in the HDEQs, 
we assume that 
the change in energy of an electron at a scattering process  
$\Delta \ve \ll \pi \kB \Te$ 
($\Delta \ve \sim e E_{\rm vh} \ell$ 
as shown in our previous paper\cite{Akera00})
and that the lattice temperature $T_{\rm L}=0$, and neglect 
the energy dependence of the density of states 
and the spin splitting.

\section{Hydrodynamic Equations}

\subsection{Macroscopic Variables}

Macroscopic variables in the present system are 
the electron temperature $\Te(\vr,t)$,
the chemical potential $\mu(\vr,t)$, and the electric field $\vE(\vr,t)$,  
with $\vr=(x,y)$. 
We introduce a reference energy $\em(\vr,t)$ defined by  
$\em=(\ve_{N \uparrow}+\ve_{N+1 \downarrow})/2$, 
where $\ve_{\alpha}$ is the energy of extended states
in the $\alpha$th level, 
and the indices $N$ and $N+1$ are those of the two Landau levels 
in the both sides of $\mu$. 
The deviation of $\mu$ from $\em$ is introduced and 
is denoted by $\mur =\mu -\em$. 
The electric field is given by $\vE=(E_x, E_y)=\vnabla \em /e $
where $\vnabla=(\partial/\partial x, \partial/\partial y)$ and $e>0$. 

$\Te$ and $\mur$ are intensive variables corresponding to 
the energy density $U$ and the electron density $n$, respectively,  
and are determined by 
the evolution equation of $U$ and that of $n$, 
which are derived in the next two subsections.  
The increase of $\Te$ produces the increase of $\rho_{xx}$,  
and the change of $\mur$ at high $\Te$ gives 
the deviation of $\rho_{xy}$ from the quantized value, 
as shown in eq.(\ref{eq:sxy}).

\subsection{Equation of the Charge Conservation}

The electron density  
averaged over regions within area $S$ 
satisfying $\lvh \ll S^{1/2}\ll \lc1$ is 
%====================================================
\begin{equation}
n(\vr,t)= {1 \over S} \sum_{\alpha,i,a} \faia \ .
\end{equation}
%====================================================
Here the integration over a continuous momentum $a$ 
is denoted by the summation, for simplicity. 
The drift and hopping processes give electron transfers 
between neighboring regions.
Since the change of $n$ is given only by electron transfers 
between the area $S$ and the outside through the boundary $L$, 
we have 
%====================================================
\begin{equation}
S {\partial n \over \partial t} 
=\sum_{\alpha,i \in S,a} {\partial \faia \over \partial t}
= -\sum_{i \in L} J_i
\ , 
\end{equation}
%====================================================
with $J_i$ the number of electrons going from the $i$th region 
to the outside of $S$ per unit time. 
Introducing the averaged electron flux density $\vj_n$, 
we have 
%====================================================
\begin{equation}
\sum_{i \in L} J_i = \int_L \vj_n \cdot \vn \rmd L
=\int_S \vnabla \cdot \vj_n \rmd S
\ , 
\end{equation}
%====================================================
where $\vn$ is the unit vector perpendicular to the boundary $L$ 
directed to the outside, 
and the two-dimensional Gauss theorem is used in the second equality. 
Since the variation of $\vnabla \cdot \vj_n$ within $S$ is 
negligible, we obtain the equation of the charge conservation: 
%====================================================
\begin{equation}
{\partial n \over \partial t} 
= -\vnabla \cdot \vj_n \ .
\end{equation}
%====================================================
The electron flux density $\vj_n$ is given by  
%====================================================
\begin{equation}
\vj_n =  \vj_n^{\rm d} + \vj_n^{\rm h} \ , 
\end{equation}
%====================================================
where $\vj_n^{\rm d}$ and $\vj_n^{\rm h}$ are 
the electron flux densities due to drift and hopping processes, respectively.

\subsection{Equation of the Energy Conservation}

The energy density due to the thermal activation, 
averaged over regions within area $S$ 
satisfying $\lvh \ll S^{1/2}\ll \lc1$,  is 
%====================================================
\begin{equation}
U(\vr,t)= {1 \over S} \sum_{\alpha, i,a} 
(\eaia-\mu_i) (\faia - \faia^0)\ , 
\label{eq:U0}
\end{equation}
%====================================================
where $\mu_i=\mu(\vr_i,t)$ and   
$\faia^0$ is the occupation probability in the local equilibrium at $\Te =0$: 
$\faia^0=f(\eaia-\mu_i, 0)$.  
Terms with $\eaia>\mu_i$ express electron excitations, and 
those with  $\eaia<\mu_i$ express hole excitations: 
%====================================================
\begin{eqnarray}
(&&\eaia -\mu_i) (\faia - \faia^0)  \nonumber \\
&&\ \ \ =\left\{
\begin{array}{ll}
(\eaia-\mu_i) \faia, &\eaia>\mu_i \\
(\mu_i-\eaia) (1-\faia), &\eaia<\mu_i \ . 
\end{array} \right.
\end{eqnarray}
%====================================================

$U$ is considered as a function of the electron distribution 
described by $\mur$ and $\faia$.  
In $\partial U/ \partial t$, 
terms with $\partial \mur/\partial t$ are neglected 
since $\mur$ changes much slower in time than $\faia$. 
Then $\partial U/ \partial t$ is given, 
using the Boltzmann-type equation eq.(\ref{eq:ft}), as
%====================================================
\begin{eqnarray}
{\partial U \over  \partial t}
&=&
\left({\partial U \over \partial t}\right)_{\rm d}
+\left({\partial U \over \partial t}\right)_{\rm p} \nonumber \\
&+&\left({\partial U \over \partial t}\right)_{\rm h1}
+\left({\partial U \over \partial t}\right)_{\rm h2}
+\left({\partial U \over \partial t}\right)_{\rm s}
\ .  
\end{eqnarray}
%====================================================
Here we have decomposed the term due to h-processes into two terms: 
h1 and h2 denote the contribution from hopping electrons 
and that from electrons scattered within a region, respectively.

The drift and hopping processes give energy transfers between regions. 
The transferred energy $|\eaia-\mu_i|$ 
is $\vera - \mur$ for electron excitations 
and $\mur - \vera$ for hole excitations 
with $\vera=\ve_{\alpha}-\em$. 
Although $\eaia -\em$ of the initial and final states 
in the hopping processes has fluctuations 
around $\vera$ due to the fluctuating potential, 
such fluctuations are canceled by 
those in the energy of the scattered electron 
in the calculation of the h2 term (see below). 
The s-processes also give energy transfers between regions. 
The change of $U$ due to such energy transfers is given by 
%====================================================
\begin{equation}
\left({\partial U \over \partial t}\right)_{\rm d}
+\left({\partial U \over \partial t}\right)_{\rm h1}
+\left({\partial U \over \partial t}\right)_{\rm s}
=-\vnabla \cdot \vj_U
\ , 
\end{equation}
%====================================================
following the procedure in the previous subsection.  
Here the averaged energy flux density $\vj_U$ is given by  
%====================================================
\begin{equation}
\vj_U =  \vj_U^{\rm d} + \vj_U^{\rm h} + \vj_U^{\rm s}
\ , 
\end{equation}
%====================================================
where $\vj_U^{\rm d}$, $\vj_U^{\rm h}$, and $\vj_U^{\rm s}$ are 
components due to drift, hopping, 
and scattering (without hopping) processes, respectively. 

The 2DES gains an energy from the electric field 
through the energy change $\Delta \ve$ of the scattered electron 
in the h-processes.  
Neglecting the fluctuations in $\Delta \ve$ (see above), 
$\Delta \ve=(-e) \vE \cdot \Delta\vr$ 
with $\Delta\vr$ the change in position of the hopping electron. 
By summing over all hopping processes within $S$, 
we obtain
%====================================================
\begin{equation}
\left({\partial U \over \partial t}\right)_{\rm h2}
=(-e)\vE \cdot  \vj_n^{\rm h}
\ .
\end{equation}
%====================================================

Electron-phonon scatterings give changes of energy 
within each region 
and result in the energy loss after the statistical averaging 
when the lattice temperature is lower than $\Te$: 
%====================================================
\begin{equation}
\left({\partial U \over \partial t}\right)_{\rm p}
=-P_{\rm L}
\ ,
\end{equation}
%====================================================
where $P_{\rm L}>0$ is the energy loss per unit area per unit time.  

Then we obtain 
%====================================================
\begin{equation}
{\partial U \over  \partial t}= 
- \vnabla \cdot \vj_U + (-e)\vE \cdot  \vj_n - P_{\rm L} 
\ , 
\end{equation}
%====================================================
where we have used $\vE \cdot \vj_n^{\rm d}=0$.

\subsection{Boundary Conditions} 

We impose boundary conditions at the edge of the 2DES 
in solving the HDEQs.  
In the case of the boundary between the 2DES and the depleted region,  
they are 
%====================================================
\begin{equation}
\vj_n \cdot \vn_{\rm b}=0 \ ,  \ \ 
\vj_U \cdot \vn_{\rm b}=0 \ ,
\end{equation}
%====================================================
where $\vn_{\rm b}$ is the unit vector perpendicular to the boundary. 
Another boundary condition is that 
the total current is given by the external circuit. 

\def\noteBC{{\bf [[ Note:}
The following will depend on the type of contacts, 
and we must specify it.  
(The conditions on the boundary between the 2DES and the metallic contact 
having the potential $V_c$ and the temperature $T_c$ 
are $V=V_c,  \Te =T_c$, respectively.) 
{\bf ]]}}%\noteBC

\section{Evaluation of Terms in the HDEQs}
\label{sec:terms}

\subsection{Dimensionless HDEQs}

Here we introduce dimensionless units in which 
the length and the time are scaled by $\ell$ and $\omega_c^{-1}$, 
respectively. 
Dimensionless variables are defined as 
$\tvr =\vr/\ell$, 
$\tvnabla = \ell \vnabla$, 
$\tilde t =\omega_c t$, and
$\tvE =e\ell \vE / \hbar\oc $.  
We use $\hbar\oc$ to scale energies 
and introduce 
$\tTe =\kB \Te / \hbar\oc$, 
$\tmur=\mur/\hbar\oc$, and $\teaia= \eaia/\hbar\oc$.  

In the present units the equation of the charge conservation becomes 
%====================================================
\begin{equation}
{\partial \tilde n \over \partial \tilde t} 
= -\tvnabla \cdot \tvj_n \ , 
\end{equation}
%====================================================
with $\tilde n=\ell^2 n$ and $\tvj_n=(\ell/\omega_c) \vj_n$. 
The equation of the energy conservation becomes 
%====================================================
\begin{equation}
{\partial \tilde U \over  \partial \tilde t}= 
- \tvnabla \cdot \tvj_U - \tvE \cdot  \tvj_n - \tilde{P}_L 
\ , 
\label{eq:dUdt}
\end{equation}
%====================================================
with $\tilde U=(\ell^2/ \hbar\oc)  U$, 
$\tvj_U=(\ell/ \hbar \omega_c^2 ) \vj_U$, and 
$\tilde{P}_L=(\ell^2/\hbar \omega_c^2)  P_{\rm L}$.

\subsection{Time Derivatives of Electron and Energy Densities}

The electron density is given in the local equilibrium by 
%====================================================
\begin{equation}
n(\vr,t)= 
2 \int^{\infty}_{\ve_{\rm b}} \rmd \ve \rho(\ve)  f(\ve-\mu,\Te) \ , 
\end{equation}
%====================================================
where $\rho(\ve)$ is the density of states per spin per unit area  
and $\ve_{\rm b}$ is the energy of the lowest state. 
Using the assumption that $\rho(\ve)$ is constant: 
$\rho=1/(2\pi \ell^2 \hbar\oc)$, we obtain 
$n(\vr,t)= 2 \rho (\mu-\ve_{\rm b})
= 2 \rho (\mur +\em -\ve_{\rm b})$ 
as long as $f(\ve_{\rm b}-\mu,\Te) \approx 1$. 
In the dimensionless units, we have
%====================================================
\begin{equation}
{\partial \tilde n \over  \partial \tilde t} = 
{1 \over \pi} {\partial \tmur \over  \partial \tilde t} \ .
\end{equation}
%====================================================

Similarly, the energy density is given in the local equilibrium by 
%====================================================
\begin{eqnarray}
U(\vr,t)= 2 \int^{\infty}_{-\infty} &&\rmd \ve \rho(\ve) (\ve-\mu) 
\nonumber \\
\times&&[f(\ve-\mu,\Te) - f (\ve-\mu,0)]\ . 
\end{eqnarray}
%====================================================
Since $\rho$ is assumed to be constant, 
$\tilde{U}(\vr,t)$ does not depend on $\tmur$, and we have 
%====================================================
\begin{equation}
{\partial \tilde U \over  \partial \tilde t} = 
{\partial \tilde U \over  \partial \tTe}{\partial \tTe \over  \partial \tilde t}
= \tCe {\partial \tTe \over  \partial \tilde t} \ , 
\label{eq:Ce}
\end{equation}
%====================================================
with the dimensionless specific heat $\tCe={\pi\over 3}\tTe$. 

\def\noteCe{{\bf [[ Note:} $U$ is given by 
%====================================================
\begin{equation}
U(\vr,t)= 2 \rho
\int^{\infty}_{-\infty} \rmd \ve \  \ve 
[f(\ve,\Te) - f (\ve,0)]\ . 
\end{equation}
%====================================================
In the dimensionless units, it becomes 
%====================================================
\begin{equation}
\tilde{U}(\vr,t)= {1 \over \pi} 
\int^{\infty}_{-\infty} \rmd \tve \  \tve 
[f(\tve,\tTe) - f (\tve,0)]\ , 
\end{equation}
%====================================================
with $f(\tve, \tTe) =1/[\exp(\tve/\tTe) +1]$. Then $\tCe$ is given by 
%====================================================
\begin{equation}
\tCe ={1 \over \pi} 
\int^{\infty}_{-\infty} \rmd \tve \  \tve {\partial f(\tve,\tTe)  \over \partial \tTe}
={1 \over \pi} 
\int^{\infty}_{-\infty} \rmd \tve \  \tve f(1-f) {\tve \over \tTe^2}
={\tTe \over \pi} 
\int^{\infty}_{-\infty} \rmd x x^2 {e^x \over (e^x +1)^2 }
={\tTe \over \pi}{\pi^2 \over 3}
 \ . 
\end{equation}
%====================================================
{\bf ]]}}%\noteCe

\subsection{Energy Loss}

The energy loss due to electron-phonon scatterings is given by 
%====================================================
\begin{equation}
\tilde{P}_L={1 \over \tS}\sum_{\alpha,i,a,b} 
(\teaia-\teaib)\faia (1-\faib) \Wiaib  \nonumber \\
\ ,
\end{equation}
%====================================================
where $\tS=S/\ell^2$ and 
$\Wiaib$ is the number during the time $\oc^{-1}$
of electron-phonon scatterings, 
in which an electron makes a transition $\alpha i a \rightarrow \alpha i b$, 
and a phonon is created or destructed depending on the sign of 
$\eaia-\eaib$.

We evaluate $\tilde{P}_L$ 
when the lattice temperature $T_L=0$. 
Here we assume that $|\eaia-\eaib| \ll \pi \kB \Te$, 
which is approximately satisfied since  
$|\eaia-\eaib| \sim e E_{\rm vh} \ell \sim 0.1 \hbar\oc$ 
and $\pi \kB \Te \sim \hbar\oc$ at $\Te = \Tc1$ from 
eq.(\ref{eq:Tc1}). 
Using this assumption, we have  
$\faia (1-\faib) \approx \faia (1-\faia) 
= -\kB \Te \partial f(\eaia -\mu,0)/\partial \ve$,   
and then obtain 
%====================================================
\begin{equation}
\tilde{P}_L=\Cp \tTe\ ,
\end{equation}
%====================================================
with 
%====================================================
\begin{eqnarray}
\Cp={1 \over \tS}
&\sum_{\alpha,i,a}& (-1) {\partial f \over \partial \ve}(\eaia -\mu,0) 
\nonumber \\
\times&\sum_{b}& (\eaia-\eaib) \Wiaib \ .
\end{eqnarray}
%====================================================
We here neglect the weak $\alpha i a$ dependence of $\Wiaib$
($\Wiaib$ depends more strongly on $\eaib-\eaia$  
through the overlap integral between the states). 
Using also the assumption of no energy dependence of the density of states, 
$\Cp$ becomes independent of $\Te$ and $\mur$, and 
is given by
%====================================================
\begin{equation}
\Cp={1 \over \pi}\sum_{b} (\teaia-\teaib) \Wiaib 
\ .
\end{equation}
%====================================================

\subsection{Drift Components of Electron and Energy Fluxes}
\label{sec:drift}

The electron flux density, averaged in a macroscopic scale, 
due to drift processes, is given by 
%====================================================
\begin{equation}
\tvj_n^{\rm d}=\sum_{\alpha} \tvj_{n \alpha}^{\rm d}\  ,
\end{equation}
%====================================================
where $\tvj_{n \alpha}^{\rm d}$ is the averaged electron flux density 
carried by electrons in the level $\alpha$. 
The energy flux density, on the other hand, is given by
%====================================================
\begin{eqnarray}
\tvj_U^{\rm d}&=&
\sum_{\alpha (\ve_{\alpha} >\mu)} (\tvera -\tmur)
\tvj_{n \alpha}^{\rm d} 
\nonumber \\
&+& \sum_{\alpha (\ve_{\alpha} <\mu)} (\tmur -\tvera)
( \tvj_{n \alpha 1}^{\rm d} - \tvj_{n \alpha}^{\rm d})
\  ,
\end{eqnarray}
%====================================================
where $\tvj_{n \alpha 1}^{\rm d}$ is 
the value of $\tvj_{n \alpha}^{\rm d}$ 
when the level $\alpha$ is completely filled by electrons. 

Since $\tvj_{n \alpha}^{\rm d}$ is an averaged flux density, 
it is only from extended states and is given by 
%====================================================
\begin{eqnarray}
\tvj_{n \alpha}^{\rm d}
&=& f_{\alpha} \tvj_{n \alpha 1}^{\rm d}
\ , \\
f_{\alpha}&=& f(\vera-\mur,\Te) \ .
\end{eqnarray}
%====================================================
It was shown by MacDonald {\it et al.}~\cite{MacDonald83} 
that the local electron flux density of the filled Landau level 
due to the drift motion in the local electric field $\vE_{\rm loc}$ 
is independent of the Landau index 
(therefore denoted as $\vj_{n1}^{\rm d,loc}$) 
when $\vE_{\rm loc}$ varies in the scale $\gg \ell$, 
and that 
$\vj_{n1}^{\rm d,loc}= (e/h)\vE_{\rm loc} \times \vez$, 
with $\vez$ the unit vector along $z$ 
($\vB=B\vez$ with $B>0$). 
In the dimensionless units, 
$\tvj_{n1}^{\rm d,loc}= {1 \over 2\pi}\tvE_{\rm loc} \times \vez$. 
By averaging this equation over $S$, we obtain  
%====================================================
\begin{equation}
\tvj_{n1}^{\rm d} 
= {\textstyle {1 \over 2\pi}} \tvE \times \vez  
= {\textstyle {1 \over 2\pi}} (\tE_y, -\tE_x)
\ ,
\end{equation}
%====================================================
and $\tvj_{n \alpha 1}^{\rm d}=\tvj_{n1}^{\rm d}$. 
The electron flux density $\tvj_n^{\rm d}$ is then given by  
%====================================================
\begin{eqnarray}
\tvj_n^{\rm d}&=&  - \tsxy (\tE_y, -\tE_x)
\ , \\
\tsxy &=& \sxy / (e^2/\hbar) 
= -{\textstyle {1 \over 2\pi}} \sum_{\alpha} f_{\alpha}
\ .
\label{eq:sxy}
\end{eqnarray}
%====================================================
When $\mur=0$, $\tsxy= - {N \over \pi}$ 
with $N$ the number of filled Landau levels. 

The energy flux density, on the other hand, is given by 
%====================================================
\begin{eqnarray}
\tvj_U^{\rm d}&=&  \tuxy (\tE_y, -\tE_x)
\ ,  \\
\tuxy&=& {\textstyle {1 \over 2\pi}} 
\sum_{\alpha (\ve_{\alpha} >\mu)} (\tvera -\tmur)
f_{\alpha}
\nonumber \\
&+& {\textstyle {1 \over 2\pi}} 
\sum_{\alpha (\ve_{\alpha} <\mu)} (\tmur -\tvera)
( 1 - f_{\alpha} )
\  .
\end{eqnarray}
%====================================================
When we consider only the two Landau levels near $\mu$, 
assume $\mur=0$, and use the assumption of no spin splitting, 
we obtain $\tuxy= {\textstyle {1 \over \pi}} f_1$
with 
%====================================================
\begin{equation}
f_1  =\left[  \exp\left( 1/ 2 \tTe \right)  +1 \right]^{-1} 
\ .
\end{equation}
%====================================================

\def\noteVector{{\bf [[ Note:}
We have  
$\tvnabla \cdot \tvj_U^{\rm d}= (\tvE \times \vez) \cdot \tvnabla \tuxy$ 
in the HDEQ. 
This is derived by using 
$\vnabla \cdot g \vj =\vj \cdot \vnabla g +g \vnabla \cdot \vj$ 
and $\tvnabla \cdot (\tvE \times \vez) =0$. 
{\bf ]]}}%\noteVector

\subsection{Hopping Components of Electron and Energy Fluxes}

The electron and energy flux densities, averaged in a macroscopic scale, 
due to electron-electron scatterings (h-processes) are given by 
%====================================================
\begin{eqnarray}
\hskip -1cm
\tvj_n^{\rm h}&=&{1 \over \tS} \sum_{\alpha,\beta,i,k,l,a,b,c,d} 
(\tvr_l -\tvr_k)  \Jiaibkcld,
\\
\hskip -1cm
\tvj_U^{\rm h}&=&{1 \over \tS} \sum_{\alpha,\beta,i,k,l,a,b,c,d} 
(\tverb -\tmur)(\tvr_l -\tvr_k)  \Jiaibkcld,
\end{eqnarray}
%====================================================
where $\tS=S/\ell^2$ and 
%====================================================
\begin{equation}
\Jiaibkcld=\faia  (1-\faib) \fbkc(1-\fbld) \Wiaibkcld
\ ,
\end{equation}
%====================================================
with $\Wiaibkcld$ 
the number of scatterings during the period $\omega_c^{-1}$. 
Regions $k$ and $l$ are the nearest neighbors. 
For hole excitations ($\tverb -\tmur <0$), 
the above formula of $\tvj_U^{\rm h}$ is consistent with 
the picture that each hole excitation carries 
the energy of $\tmur -\tverb$, 
since the hole flux is opposite in sign to the electron flux. 

We rewrite $\tvj_n^{\rm h}$ as
%====================================================
\begin{equation}
\tvj_n^{\rm h}=
{1 \over 2\tS} \sum_{\alpha,\beta,i,k,l,a,b,c,d} 
(\tvr_l -\tvr_k)  (\Jiaibkcld-\Jibialdkc) \  ,
\end{equation}
%====================================================
and $\tvj_U^{\rm h}$ similarly, 
where changes of labels are made in the term with $\Jibialdkc$: 
$a\leftrightarrow b$ and $kc\leftrightarrow ld$. 
Since the two terms are forward and backward transitions 
among the same set of states, 
$\Jiaibkcld-\Jibialdkc$ is 
the net number (during $\omega_c^{-1}$) of electrons 
which hop from $c$ in $k$ to $d$ in $l$, and 
is induced by $\vnabla \mu$ and $\vnabla \Te$. 
In the first order of $\vnabla \mu$ and $\vnabla \Te$, 
$\Jiaibkcld-\Jibialdkc$ is given by
%====================================================
\begin{eqnarray}
&&\Jiaibkcld-\Jibialdkc \nonumber \\
&&=
[- \tTe^{-1}(\tmu_l -\tmu_k)
+ (\tverb -\tmur) (\tT_l^{-1}- \tT_k^{-1}) ]\Jiaibkcld \ ,\nonumber \\
&&=
(\tvr_l -\tvr_k) \cdot [- \tTe^{-1} \tvnabla \tmu
+ (\tverb -\tmur)  \tvnabla \tTe^{-1}]\Jiaibkcld \ ,\nonumber \\
\end{eqnarray}
%====================================================
where $\tmu_k=\mu(\vr_k,t)/\hbar\oc$, $\tT_k =\tTe(\vr_k,t)$, and 
$\Jiaibkcld$ in the right should be evaluated 
in the absence of $\vnabla \mu$ and $\vnabla \Te$. 
Higher-order terms are negligible when 
$|\vnabla\mu| \lvh / \kB \Te  \ll 1$ and 
$|\vnabla\Te| \lvh / 2\Te \tTe \ll 1$. 
These conditions are approximately satisfied 
at $E=\Ec1$ and $\Te=\Tc1$:     
since the estimated value of $\Ec1$ in eq.(\ref{eq:Ec1}) 
and that of $\Tc1$ in eq.(\ref{eq:Tc1}) give 
$|\vnabla\mu| \lvh / \kB \Te  \sim 0.1$ and 
$|\vnabla\Te| \lvh / 2\Te \tTe \sim 0.1$. 
We also have made an approximation that 
$\ebkc-\mu_k, \ebld-\mu_l \approx  \verb - \mur$. 

\def\noteJJ{{\bf [[ Note:}
The derivation is the following. 
%====================================================
\begin{eqnarray}
\Jiaibkcld-\Jibialdkc
&=&
[\faia  (1-\faib) \fbkc(1-\fbld)  - \faib  (1-\faia) \fbld(1-\fbkc) ]
\Wiaibkcld   \nonumber \\
&=& \faia  (1-\faib) \fbkc(1-\fbld) (1- e^{\Delta z}) 
\Wiaibkcld \ ,
\end{eqnarray}
%====================================================
with 
%====================================================
\begin{eqnarray}
\Delta z 
&=& \beta_i(\eaia-\mu_i) - \beta_i(\eaib-\mu_i) 
+\beta_k(\ebkc-\mu_k) - \beta_l(\ebld-\mu_l)  \nonumber \\
&=& \beta_i(\eaia-\eaib) 
+\beta_i (\ebkc-\ebld)
-\beta_i(\mu_k -\mu_l)
+(\beta_k- \beta_i)(\ebkc-\mu_k) 
- (\beta_l- \beta_i)(\ebld-\mu_l) \nonumber \\
&\approx& -\beta_i(\mu_k -\mu_l) 
+(\verb -\mur)(\beta_k- \beta_l) \ .
\end{eqnarray}
%====================================================
In the last equality we have used 
$(\eaia-\eaib) +(\ebkc-\ebld)=0$ and 
$\ebkc-\mu_k \approx \ebld-\mu_l \approx  \verb -\mur$.   
We make an approximation:
$1- e^{\Delta z} = - \Delta z $, 
since, when $\tilde E=\tEc1$ and $\tTe=\tTc1$, 
%====================================================
\begin{eqnarray}
|\mu_k -\mu_l|/\kB \Te  
&\leq& eE\lvh / 0.3 \hbar\oc
=\tilde E (\lvh/\ell)/0.3 
\approx 0.003 \times 10 / 0.3 = 0.1 \ ,
\\
|\beta_k -\beta_l|\hbar\oc/2 
&=&{|T_k - T_l| \over \kB \Te^2} \hbar\oc/2
={|T_k - T_l| \over 2\Te \tTe} 
\leq {\lvh \over 2\lc1 \times 0.3} \approx 0.1 \ .
\end{eqnarray}
%====================================================
Then we have
$\Jiaibkcld-\Jibialdkc =\Jiaibkcld
[\beta_i(\mu_k -\mu_l) -(\verb -\mur)(\beta_k- \beta_l) ] $. 
{\bf ]]}}%\noteJJ

We then have 
%====================================================
\begin{eqnarray}
\tvj_n^{\rm h}&=&
{1 \over 2\tS} \sum_{\alpha,\beta,k,l} 
(\tvr_l -\tvr_k)  
\nonumber \\
&\times& (\tvr_l -\tvr_k) \cdot 
[- \tTe^{-1} \tvnabla \tmu
+ (\tverb -\tmur) \tvnabla \tTe^{-1}] 
\nonumber \\
&\times&\sum_{i,a,b,c,d} \Jiaibkcld 
 \  .
\end{eqnarray}
%====================================================
Here we make the integration over $\theta_{lk}$, 
the polar coordinate of 
$\vr_l - \vr_k=r_{lk} \exp(i\theta_{lk})=(x_{lk},y_{lk})$. 
The dependence on $\vr_l - \vr_k$ appears 
through $\Jiaibkcld$ and $\tvr_l -\tvr_k$, but 
the $\theta_{lk}$ dependence appears only through $\tvr_l -\tvr_k$,  
since $\sum_{i,a,b,c,d} \Jiaibkcld$ has no dependence on $\theta_{lk}$. 
We assume the uniform distribution of $\theta_{lk}$,  
in which we have 
$\left< x_{lk} y_{lk} \right>=0$ and 
$\left< (x_{lk})^2 \right>=\left< (y_{lk})^2 \right>= (r_{lk})^2 /2$, 
with 
$\left<\cdots\right>\equiv 
{1\over 2\pi} \int^{2\pi}_0 \cdots \rmd \theta_{lk}$.
We also make an approximation that $r_{lk} \approx \lvh$. 

\def\noteAve{{\bf [[ Note:}
We then obtain, using $\vd \equiv \vr_l - \vr_k=(d_1,d_2)$, 
%====================================================
\begin{eqnarray}
\left< \td_n (\tvd \!\cdot\! \tvnabla \tmu) \right>
= \sum_{m=1,2} \left<\td_n \td_m \right>\! \tnabla_m \tmu 
= \left<\td_n^2 \right> \tnabla_n \tmu
= {\tlvh^2 \over 2} \tnabla_n \tmu \ , \\
\tvj_{n}^{\rm h}=
{1 \over 2\tS} \sum_{\alpha,\beta,k,l} {\tlvh^2 \over 2} 
\left[ - \tTe^{-1}\tvnabla \tmu 
+(\tverb - \tmur) \tvnabla \tTe^{-1} \right]
\sum_{i,a,b,c,d} \Jiaibkcld
\  .
\end{eqnarray}
%====================================================
{\bf ]]}}%\noteAve

Then we obtain the formulas of $\tvj_{n}^{\rm h}$ and $\tvj_{U}^{\rm h}$: 
%====================================================
\begin{eqnarray}
\tvj_{n}^{\rm h}&=& L_{n\mu} \left(- \tTe^{-1}\tvnabla \tmu\right) 
                                  +  L_{nT} \tvnabla \tTe^{-1}  \ , 
\\
\tvj_{U}^{\rm h}&=& L_{U\mu} \left(- \tTe^{-1}\tvnabla \tmu\right) 
                                  +  L_{UT} \tvnabla \tTe^{-1}   \ ,
\end{eqnarray}
%====================================================
with 
%====================================================
\begin{eqnarray}
L_{n\mu} = L_0 \ , \ \ 
L_{nT} =L_{U\mu}=L_1 \ , \ \ 
L_{UT} =L_2 \ , \ \ \ \ \ \\
L_m ={\lvh^2 \over 4S} \sum_{\alpha, \beta,i,k,l,a,b,c,d} 
(\tverb - \tmur)^m \Jiaibkcld \ , \ \ \ \ \ \nonumber \\
(m=0,1,2) \ , \ \ \ \ \ 
\end{eqnarray}
%====================================================
where 
$\vnabla \mu=e\vE +\vnabla \mur \approx e\vE$ and 
in dimensionless units $\tvnabla \tmu \approx \tvE$.  

Here we assume $|\eaia-\eaib| \ll \pi \kB \Te$.  
Then we have 
$\faia (1-\faib) \approx \faia (1-\faia) 
= -\kB \Te \partial f(\eaia-\mu,\Te)/\partial \ve$ 
and $\fbkc, \fbld \approx f_{\beta}$. 
We neglect weak dependences of $\Wiaibkcld$ 
on $\alpha, i, a$, and $\beta$, for simplicity. 
We then obtain, by performing the summation over $\alpha,i,a$,  
%====================================================
\begin{equation}
L_m ={\textstyle {1 \over 4}} \Ch \tTe 
\sum_{\beta} (\tverb - \tmur)^m f_{\beta}(1-f_{\beta})  \ .
\end{equation}
%====================================================
Here 
%====================================================
\begin{equation}
\Ch = 
{\tlvh^2 \over \pi } \sum_{k,l,b,c,d} \Wiaibkcld \ ,
\end{equation}
%====================================================
with $\tlvh \equiv \lvh/\ell$, has no dependence on $\Te$ and $\mur$.  

When we consider only the two Landau levels near $\mu$, 
assume $\mur=0$, and use the assumption of no spin splitting, 
we obtain 
$L_{n\mu} =\Ch \tTe f_1(1-f_1)$, 
$L_{nT} =L_{U\mu}=0$,
and 
$L_{UT} = {\textstyle {1 \over 4}} \Ch \tTe  f_1(1-f_1)$. 
In the present case $L_{nT}=0$ because 
energy levels of extended states are symmetric with respect to $\mu$ 
and electron fluxes induced by $\vnabla \Te$  
are canceled between the upper and lower Landau levels,  
and $L_{U\mu}=0$ because 
electrons and holes move in opposite directions by hopping processes 
in the presence of $\vE$, in contrast with drift processes,   
and energy fluxes due to such hopping processes are 
canceled between the upper and lower Landau levels. 
The electron and energy flux densities in this case are given 
in terms of the electrical conductivity $\sxx$ and 
the thermal conductivity $\kappa_{\rm h}$ as 
%====================================================
\begin{eqnarray}
\tvj_{n}^{\rm h}&=& - \tsxx \tvE  \ , \\
\tvj_{U}^{\rm h}&=& - \tk_{\rm h} \tvnabla \tTe  \ ,
\end{eqnarray}
%====================================================
with 
%====================================================
\begin{eqnarray}
\tsxx           &=&\sxx/(e^2/\hbar) =\Ch f_1(1-f_1) \ , \\
\tk_{\rm h}&=&\kappa_{\rm h}/(\kB \oc) 
={\textstyle {1 \over 4}} \tTe^{-1} \tsxx  \ .
\end{eqnarray}
%====================================================

\def\noteCh{{\bf [[ Note:}
In estimating $\Ch$, we only consider terms 
in which one of $k,l$ coincides with $i$, 
since the terms with larger distances between $i$ and $k,l$ 
can be neglected 
[the Coulomb matrix element, 
which is approximately the interaction between a point charge and a dipole, 
decreases with the distance $d$ as $d^{-2}$ 
and the transition rate, $\Wiaibkcld$, as $d^{-4}$.  
The summation becomes 
$\sum_d 2 \pi d \times d^{-4} = 2 \pi\sum_d   d^{-3}$, 
and the contributions from larger $d$'s are not large 
($\sum_{d=1}^{d=\infty}   d^{-3} \approx 1.2$ and 
$\sum_{d=2}^{d=\infty}   d^{-3} \approx 0.2$)].
Using $N_c$ the coordination number, 
we have
%====================================================
\begin{equation}
\Ch = {\tlvh^2 \over \pi} 2N_c
\sum_{b,c,d} \Wiaibkcid \ .
\end{equation}
%====================================================
where $k$ is the nearest neighbor of $i$.  
The above formula of $\Ch$ is the same as 
that of the coefficient of $P_{\rm G}$ given in 
our previous paper.~\cite{Akera00}
{\bf ]]}}%\noteCh

\subsection{S-process Component of Energy Flux}

The energy flux density, averaged in a macroscopic scale, 
due to electron-electron scatterings without hoppings, is 
%====================================================
\begin{equation}
\tvj_U^{\rm s}
={1 \over 2\tS} \sum_{\alpha,\beta,i,k,a,b,c,d} 
(\teaib-\teaia)  (\tvr_i - \tvr_k ) \Jiaibkckd 
 \ ,
\end{equation}
%====================================================
where 
%====================================================
\begin{equation}
\Jiaibkckd=\faia  (1-\faib) \fbkc(1-\fbkd)  \Wiaibkckd \ .
\end{equation}
%====================================================
The factor $1/2$ in the formula of $\tvj_U^{\rm s}$ corrects 
the double counting: $i=1, k=2$ and $i=2, k=1$, for example. 
Following the derivation of $\tvj_U^{\rm h}$ in the previous subsection, 
we obtain 
%====================================================
\begin{equation}
\tvj_U^{\rm s}  =  - \tk_{\rm s} \tvnabla \tTe \ ,
\end{equation}
%====================================================
when $|\vnabla \Te| \lvh / \Te \ll 1$, 
with the thermal conductivity
%====================================================
\begin{equation}
\tk_{\rm s}
= {1 \over 8\tS \tTe^2} \sum_{\alpha,\beta,i,k,a,b,c,d} 
\Jiaibkckd (\teaib-\teaia)^2  (\tvr_i - \tvr_k )^2 \ .
\end{equation}
%====================================================

\def\noteJUs{{\bf [[ Note:}
We calculate $\Jiaibkckd-\Jibiakdkc$
in the first order of $\vnabla \Te$, 
using the expression of $\faia$ in the local equilibrium. 
We write it as
%====================================================
\begin{eqnarray}
\Jiaibkckd-\Jibiakdkc
&=& [\faia  (1-\faib) \fbkc(1-\fbkd)
-\faib  (1-\faia) \fbkd(1-\fbkc)]
\Wiaibkckd   \nonumber \\
&=& \faia  (1-\faib) \fbkc(1-\fbkd) (1- e^{\Delta z}) 
\Wiaibkckd \ ,
\end{eqnarray}
%====================================================
with 
%====================================================
\begin{equation}
\Delta z = \beta_i(\eaia-\mu_i) - \beta_i(\eaib-\mu_i) 
+\beta_k(\ebkc-\mu_k) - \beta_k(\ebkd-\mu_k)
=(\beta_k -\beta_i)(\eaib-\eaia) \ .
\end{equation}
%====================================================
We make an approximation that $1- e^{\Delta z} = - \Delta z $,
assuming $\lvh |\nabla \Te| / \Te \ll 1$. 
Then we obtain 
%====================================================
\begin{equation}
\tvj_U^{\rm s} 
= {1 \over 4\tS} \sum_{\alpha,\beta,i,k,a,b,c,d} 
\Jiaibkckd (\teaib-\teaia)^2 (\tvr_k -\tvr_i) (\tT_k^{-1} -\tT_i^{-1})
 \ .
\end{equation}
%====================================================
Performing the average over the angle of $\Delta \vr= \vr_k -\vr_i$, 
we have 
%====================================================
\begin{equation}
\tvj_U^{\rm s} 
= {1 \over 8\tS} \sum_{\alpha,\beta,i,k,a,b,c,d} 
\Jiaibkckd (\teaib-\teaia)^2 (\tvr_k -\tvr_i)^2 \tvnabla \tTe^{-1}
\ .
\end{equation}
%====================================================
{\bf ]]}}%\noteJUs

Here we assume that $|\eaia-\eaib| \ll \pi \kB \Te$ and 
neglect the weak $\alpha,\beta,a,c$ dependence of $\Wiaibkckd$. 
We restrict the calculation to the case of $\mur=0$ and no spin splitting. 
Then $\tk_{\rm s}$ at $\mur=0$, 
which is denoted by $\tk_{\rm s0}$, is given by
%====================================================
\begin{equation}
\tk_{\rm s0} 
= {\tlvh^4 \over 8 \pi^2 \tS} \sum_{i,k,b,d} 
\Wiaibkckd (\teaib-\teaia)^2  (\tvr_i - \tvr_k )^2 \ .
\end{equation}
%====================================================
$\tk_{\rm s0}$ has no dependence on $\Te$. 

\def\noteKappa{{\bf [[ Note:} The derivation is the following. 
When region $i$ is a valley,  
%====================================================
\begin{equation}
\sum_{\alpha,a} [- f'(\eaia -\mu,\Te) ]  
= 2 \sum_{n=-\infty}^{\infty} 
\int^{(n+1/2)\hbar\oc}_{n\hbar\oc} \rho_r \rmd \ve 
(-1)f'(\ve,\Te)   
=  \rho_r \ ,
\end{equation}
%====================================================
and, when region $i$ is a hill, 
%====================================================
\begin{equation}
\sum_{\alpha,a} [- f'(\eaia -\mu,\Te) ]  
= 2 \sum_{n=-\infty}^{\infty} 
\int^{n\hbar\oc}_{(n-1/2)\hbar\oc} \rho_r \rmd \ve 
(-1)f'(\ve,\Te)   
=  \rho_r \ ,
\end{equation}
%====================================================
with $\rho_r=S_{\rm vh}/2\pi \ell^2 \hbar\oc$ and 
$S_{\rm vh}=2\lvh^2$. 

In estimating $\tk_{\rm s0}$, 
we limit the summation over $k$ to the nearest neighbors of $i$, 
since the terms with larger distances between $k$ and $i$ 
can be neglected 
[the Coulomb matrix element, 
which is approximately the interaction between dipoles, 
decreases with the distance $d$ as $d^{-3}$ 
and the transition rate, $\Wiaibkckd$, as $d^{-6}$
(the matrix elements are smaller by 1/10 
for $i,k={\rm valley}({\rm hill})$ 
than for $i={\rm valley}({\rm hill}),k={\rm hill}({\rm valley})$).  
The summation becomes 
$\sum_d 2 \pi d \times d^{-6} d^2= 2 \pi\sum_d   d^{-3}$, 
and the contributions from larger $d$'s are not large].
The summation over $i$ consists of two cases: 
$i={\rm valley}$ ($k={\rm hill}$)
and $i={\rm hill}$ ($k={\rm valley}$). 
Both give the same contribution. 
Then we have
%====================================================
\begin{equation}
\tk_{\rm s0} 
= {N_c \tlvh^6 \over 4 \pi^2 \tS} \sum_{i={\rm valley},b,d} 
\Wiaibkckd (\teaib-\teaia)^2    
= {N_c \tlvh^4  \over 8 \pi^2 } \sum_{b,d} 
\Wiaibkckd (\teaib-\teaia)^2 \ .
\end{equation}
%====================================================
{\bf ]]}}%\noteKappa

\subsection{Inelastic scattering time}
\label{sec:inelastic}

Here we calculate the inelastic scattering time $\tau_{\alpha i a}$ 
of state $\alpha i a$ due to electron-electron scatterings. 
Considering s-processes, 
$\tau_{\alpha i a}$ is given by 
%====================================================
\begin{equation}
{1 \over \oc \tau_{\alpha i a}}
= \sum_{\beta,k,b,c,d} (1-\faib) \fbkc(1-\fbkd) \Wiaibkckd  \ .
\end{equation}
%====================================================
Following the evaluation of $\tk_{\rm s}$,   
we assume $|\eaia-\eaib| \ll \pi \kB \Te$ and 
neglect the $\beta,c$ dependence of $\Wiaibkckd$. 
Then we obtain, when $\mur=0$,  
%====================================================
\begin{equation}
{1 \over \oc \tau_{\alpha i a}}
= {\textstyle{1 \over \pi}} \tlvh^2 \tTe (1-\faia) 
   \sum_{k, b,d} \Wiaibkckd 
\ .
\end{equation}
%====================================================
An estimation gives $1/\oc \tau_{\alpha i a}\sim 0.2$ 
for an energy $\eaia-\em \sim \hbar\oc /2$ 
at $\mur=0$, $B=$5T, and $\Te=\Tc1$ 
(the value of $\Tc1$ is given in eq.(\ref{eq:Tc1})). 
Then the level broadening $\Gamma$ is estimated to be 
$\Gamma=\hbar/\tau_{\alpha i a}\sim 0.2 \hbar\oc$.  
The inelastic scattering length defined by 
$\ell_{\rm ee}=\tau_{\alpha i a}v_{\rm vh}$ is estimated to be 
$\ell_{\rm ee} / \ell  
=(\oc \tau_{\alpha i a}) eE_{\rm vh} \ell / \hbar\oc \sim 1$.

\section{Estimation of Coefficients and Scaled HDEQs}

\subsection{Uniform Steady States at Even-Integer Filling Factors}

\begin{figure}
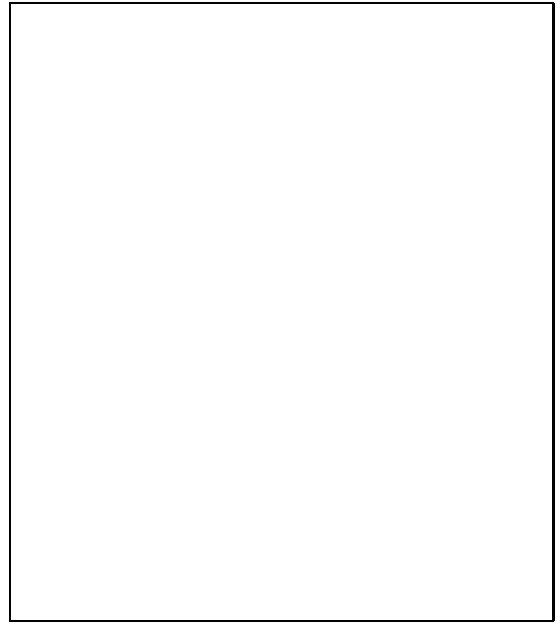

\figureheight{8cm}
\caption{ 
(a) The energy gain $\tilde P_{\rm G}=\tsxx \tilde E^2$ 
at the critical electric field $\Ec1$ 
and the loss $\tilde P_{\rm L}=\Cp \tTe$ (both divided by $\Cp$)
as a function of the electron temperature $\tTe$.  
Points of intersection give $\tTe$ in stationary states. 
(b) $\tTe$ and $\tsxx/\Ch$ in stationary states as a function of $E/\Ec1$. 
} 
\label{fig:steadystate}
\end{figure}

We here consider the simplest case of 
uniform steady states at even-integer global filling factors 
and obtain the formula of the critical electric field $\tEc1$ 
as well as the value of $\tTc1$. 
In the present case we have $\mur =0$ everywhere in the sample. 
We consider only the two Landau levels near $\mu$, 
neglect the spin splitting, and then 
use the simplified formula of $\tsxx=\Ch f_1 (1-f_1)$. 
We will give only a summary of the results, 
which we have obtained in our previous paper.~\cite{Akera00} 
The equation of the energy conservation in this case gives 
the balance of the energy gain $\tilde P_{\rm G}$ 
and the energy loss $\tilde P_{\rm L}$: 
%====================================================
\begin{equation}
\left( \tilde P_{\rm G} = \right) \tsxx \tilde E^2 
=\Cp \tTe  \left( = \tilde P_{\rm L} \right)\ .
\end{equation}
%====================================================
$\tilde P_{\rm G}$ and $\tilde P_{\rm L}$ are plotted 
as a function of $\tTe$ in Fig.~\ref{fig:steadystate}(a). 
Points of intersection of the two curves, 
$\tilde P_{\rm G}(\tTe)$ and $\tilde P_{\rm L}(\tTe)$, 
give $\tTe$ in stationary states. 
The number of points of intersection increases from one to three 
as increasing $\tE$ through a critical value.  
This critical value is $\tEc1$, and is given by 
%====================================================
\begin{equation}
\tEc1=(\Cp \gamma / \Ch)^{1/2}\ \  (\gamma = 2.2334) \ . 
\end{equation}
%==================================================== 
A bistability appears above $\tEc1$ 
(one of three in the middle corresponds to an unstable state). 
The present bistability originates from 
the existence of the activation energy $\hbar \oc /2$ of $\tilde P_{\rm G}$. 
$\tTe$ and $\tsxx$ in stationary states as a function of $E/ \Ec1$ are  
plotted in Fig.~\ref{fig:steadystate}(b). 
The value of $\tTe$ at $\tEc1$ in the high $\tTe$ branch is $\tTc1$, 
and is given by
%====================================================
\begin{equation}
\tTc1 = 0.324 \ . 
\label{eq:Tc1}
\end{equation}
%==================================================== 
%#############################################################
%#############################################################

\subsection{Estimation of Coefficients}
\label{sec:estimate}

Rough estimates of the coefficients in the HDEQs at $B=5{\rm T}$ 
have been given in our previous papers~\cite{Akera00,Akera01}
and are summarized here: 
%====================================================
\begin{equation}
\Cp = 1.0 \times 10^{-5} \ ,\ 
\Ch = 2.5 \ ,\  
\tk_{\rm s0} = 0.016   \ . 
\end{equation}
%====================================================
The ratio of the two components of the thermal conductivities, 
$\tk_{\rm h}$ and $\tk_{\rm s}$, is then estimated to be 
%====================================================
\begin{equation}
\tk_{\rm h}/\tk_{\rm s}
={\textstyle {1 \over 4}} \Ch f_1 (1-f_1) \tTe^{-1} / \tk_{\rm s0} 
\approx 17  \ ,
\end{equation}
%====================================================
when $\mur =0$ and $\Te =T_{c1}$. 
The thermal conductivity due to hopping processes is much larger 
than that due to electron-electron scatterings without hoppings. 
In the following we neglect $\tk_{\rm s}$.

The above estimates of $\Cp$ and $\Ch$ give 
theoretical estimates of $\tEc1$ and $\tsxxs={1\over 4}\Ch$, 
the saturation value of $\tsxx$ at large $\Te$ 
(only the two Landau levels near $\mu$ are considered). 
The theoretical estimates 
and the observed values~\cite{Ebert83,Komiyama85} are as follows. 
%====================================================
\begin{eqnarray}
{\rm Theory}(B=5.0{\rm T}): 
\tEc1&= 3 \times 10^{-3}  \ ,\ \tsxxs&= 0.6 \ , \ \ \ \ \ \label{eq:Ec1} \\
{\rm Ref.~}\citen{Ebert83}(B=4.7{\rm T}): 
\tEc1&= 9 \times 10^{-3}  \ ,\ \tsxxs&= 0.02 \ , \ \ \ \ \ \\
{\rm Ref.~}\citen{Komiyama85}(B=3.8{\rm T}): 
\tEc1&= 8 \times 10^{-3}  \ ,\ \tsxxs&= 0.02 \ . \ \ \ \ \ 
\end{eqnarray}
%====================================================
The discrepancies between the theory and the experiments will be 
within the limitation of accuracy in the theory, 
since a large ambiguity in the estimation of $\lvh/\ell$ 
produces large errors in the estimation of $\Cp$ and $\Ch$.  
It may be convenient to consider $\Cp$ and $\Ch$ 
as adjustable parameters 
when spatio-temporal evolutions are compared between 
the theory and the experiment.

\subsection{Scaled HDEQs}
\label{sec:scaledHDEQs}

The energy gain and the energy loss in the HDEQs 
are very small in magnitude in the units we have used: 
$\tilde P_{\rm G} \sim \tsxx \tEc1^2 \sim 10^{-5}$ 
and $\tilde P_{\rm L} \sim \Cp \tTc1 \sim 10^{-5}$ 
according to the estimation in the previous subsection. 
Therefore we switch the units to the ones in which 
the length and the time are scaled by $\lc1$ and $\tc1$, respectively,
defined in \S \ref{sec:scales}.   
Scaled variables are defined as 
$\check{\vr}= \vr/ \lc1 $, 
$\cvnabla= \lc1 \vnabla$, 
$\check t= t/ \tc1$, and
$\cvE =e\lc1 \vE /\hbar\oc$. 
Ratios between the present scale and the previous scale are  
$\lc1/\ell =(\tEc1)^{-1}$ and 
$\tc1\omega_c =(\tEc1)^{-2}$. 
Since $\cvE =\tvE / \tEc1$, 
the energy gain and loss become of the order of unity in the present units. 

When $\mur=0$ and 
only the two Landau levels near $\mu$ are considered 
for hopping processes,  
the HDEQs in the new units 
become 
%====================================================
\begin{eqnarray}
{1 \over \pi} {\partial \tmur \over  \partial \check t}
&=& -\cvnabla \cdot \cvj_n \ , 
\label{eq:nt} \\
\tCe {\partial \tTe \over  \partial \check t} 
&=& 
-  \cvnabla \cdot \cvj_U
+ \tsxx \cE^2 
- \Ch \gamma^{-1} \tTe    
 \ , 
\label{eq:Ut}
\end{eqnarray}
%====================================================
with 
%====================================================
\begin{eqnarray}
\cvj_n &=& (\lc1/\oc)\vj_n= - \tsxx \cvE -\tsxy (\cvE \times \vez)
 \ , \ \ \ \ \ 
 \label{eq:jn} \\
\cvj_U &=& (\lc1/\hbar \oc^2) \vj_U
= - \tk_{\rm h} \cvnabla   \tTe   
+ \tuxy (\cvE \times \vez)
 \ . \ \ \ \ \ 
\label{eq:jU}
\end{eqnarray}
%====================================================
The scaled HDEQs show that the typical length and time scales 
of variations of $\Te$ and $\mur$ are $\lc1$ and $\tc1$, respectively.

\section{Current Distribution at Even-Integer Filling Factors}

In this section we illustrate how to apply the HDEQs to 
the calculation of spatial variations of $\Te$. 
We consider, as an example, 
the current distribution in a steady state 
in the higher-$\Te$ branch in the middle of a long sample, 
in which spatial variations occur only perpendicular to the current. 
We restrict our calculation to the simple case where 
the global filling factor is $2N$ ($N=1,2,\cdots$) and  
$\Te$ is low enough that 
only the two Landau levels near $\mu$ contribute to 
$\vj_n^{\rm h}$,  $\vj_U^{\rm h}$, and $\vj_U^{\rm d}$. 

\def\noteExp{{\bf [[ Note:}
The experiment which confirmed the theory by MacDonald et al.: 
P.~F.~Fontein, J.~A.~Kleinen, P.~Hendriks, F.~A.~P.~Blom, J.~H.~
Wolter, H.~G.~M.~Lochs, F.~A.~J.~M.~Driessen, L.~J.~Giling, 
and C.~W.~J.~Beenakker: 
Phys.~Rev. B {\bf 43} (1991) 12090. 

Experimental suggestions that $j_x$ is uniform at the breakdown  
from the observations that the critical current for the breakdown of QHE 
is proportional to the sample width: 
S.~Kawaji, K.~Hirakawa and M.~Nagata: Physica B {\bf 184} (1993) 17;  
A.~Boisen, P.~B{\o}ggild, A.~Kristensen and 
P.~E.~Lindelof: Phys.~Rev. B {\bf 50} (1994) 1957.
{\bf ]]}}%\noteExp

\subsection{Electrostatics and Estimation of $\mur$}

In this subsection we derive the equation relating $\vE$ and $\mur$.  
We show here that $\tmur \sim (\tEc1)^2 \sim 10^{-5}$. 

We define the two-dimensional local charge density 
by $\sigma_{\rm loc} (\vr) =-e (n_{\rm loc}-n_0)$ 
with $n_{\rm loc}$ the local electron density and 
$n_0$ the uniform electron density corresponding to 
the global filling factor $2N$. 
The local charge density is divided into two components:
$\sigma_{\rm loc}=\sigma +\sigma_{\rm sr}$ with 
$\sigma$ long-ranged variations due to the Hall polarization 
and $\sigma_{\rm sr}$ short-ranged fluctuations induced 
by density fluctuations of ionized donors. 
The local electric field $\vE_{\rm loc}$ is 
also divided into two components:  
$\vE_{\rm loc}=\vE + \vE_{\rm sr}$ 
with $\vE$ the macroscopic electric field  
consisting of the Hall field induced by $\sigma$
and the uniform field along $x$ 
and $\vE_{\rm sr}$ the short-ranged fluctuations  
due to ionized donors and $\sigma_{\rm sr}$. 
The long-ranged variations given by $\sigma$ and $\vE$ are 
assumed to occur only along $y$ ($0<y<W$). 

\def\noteChd{{\bf [[ Note:}
The charge density $\sigma$ is given by
%====================================================
\begin{equation}
\sigma = 
 - {N e^2 \over \pi \hbar\oc} {\partial E_y \over \partial y} 
 - 2e \rho \mur  \ . 
\end{equation}
%====================================================
{\bf ]]}}%\noteChd

The dimensionless charge density 
$\tilde \sigma=\sigma \ell^2/e$ is given by 
%====================================================
\begin{equation}
\tilde \sigma 
=  - {N \over \pi } {\partial \tE_y \over \partial \tilde y}
-{\textstyle {1 \over \pi}} \tmur \ .
\label{eq:sigma}
\end{equation}
%====================================================
The first term of $\tilde \sigma$ was given by 
MacDonald {\it et al.}~\cite{MacDonald83}
It is the charge density in the presence of 
spatial modulations of the electric field 
when the lower $N$ Landau levels are fully occupied and 
the other higher Landau levels are empty, and 
it comes from the polarization of wave functions of the occupied states. 
The second term is the charge density 
due to the occupation of electrons or holes,  
which is described by $\mur$ when averaged in a macroscopic scale. 

The charge density $\tilde \sigma$ is also given by 
the Gauss law relating $\tilde \sigma$ and $\tE_y$: 
%====================================================
\begin{equation}
\tE_y(\tilde y)  ={e^2/\epsilon \ell \over \hbar\oc }
 \int_0^{W/\ell} \rmd \tilde y' 
 {2 \tilde \sigma(\tilde y') {\rm sgn}(\tilde y-\tilde y') \over 
[(\tilde y-\tilde y')^2 + \tilde z_{\rm c}^2]^{1/2} }
 \ , 
\label{eq:Ey}
\end{equation}
%====================================================
where $\epsilon$ is the dielectric constant and 
$\ell \tilde z_{\rm c}$ is the cutoff length corresponding to 
the finite thickness of the electron density perpendicular to the layer. 
Here 
${\rm sgn}(y)=1$ when $y>0$ and ${\rm sgn}(y)=-1$ when $y<0$.

\def\noteGauss{{\bf [[ Note:}
Applying the Gauss law to 
a line charge of density $\sigma dy$ 
at the center of a cylinder of radius $r$,  
the electric field $dE$ on the surface of the cylinder 
satisfy $2\pi r dE= 4\pi \sigma dy/\epsilon$,  
giving $dE= 2 \sigma dy /\epsilon r$. 
Integrating this over $y$ within the sample, we obtain
%====================================================
\begin{equation}
E_y(y)=
\int_0^y \rmd y'
{2 \sigma(y') \over \epsilon[(y-y')^2 + z_{\rm c}^2]^{1/2} }
- \int_y^{W} \rmd y' 
{2 \sigma(y') \over \epsilon[(y-y')^2 + z_{\rm c}^2]^{1/2} }
 \ .
\end{equation}
%====================================================
Effects of screening by the gate electrode or the surface charges 
are not taken into account. 
In the presence of such screening, 
$\sigma$ for a given $\vE$ is larger but 
its order of magnitude is the same. 
{\bf ]]}}%\noteGauss

Now we estimate the order of magnitude of $\tmur$. 
From eq.(\ref{eq:Ey}), we have 
%====================================================
\begin{eqnarray}
{\partial \tE_y \over \partial \tilde y} 
&=&{e^2/\epsilon \ell \over \hbar\oc }
\left[
 {4 \tilde \sigma(\tilde y) \over \tilde z_{\rm c}} 
- \int_0^{W /\ell} \rmd \tilde y' 
{2 \tilde \sigma(\tilde y') |\tilde y-\tilde y'| \over 
[(\tilde y-\tilde y')^2 + \tilde z_{\rm c}^2]^{3/2} } \right] 
\nonumber \\
&\sim &
{e^2/\epsilon \ell \over \hbar\oc }
{\tilde \sigma(\tilde y) \over \tilde z_{\rm c}}
\ ,
\end{eqnarray}
%====================================================
where we have assumed that $\tilde \sigma(\tilde y)$ does not change 
significantly in the scale of $\tilde z_{\rm c}$. 
Then we have  
$\tilde \sigma (\tilde y) \sim \partial \tE_y / \partial \tilde y$ 
since $e^2/\epsilon \ell \sim \hbar\oc$ and $\tilde z_{\rm c} \sim 1$. 
Combining this with eq.(\ref{eq:sigma}), we obtain 
$\tmur \sim \partial \tE_y / \partial \tilde y 
\sim \tEc1 \ell /\lc1 = (\tEc1)^2 \sim 10^{-5}$.

\subsection{Macroscopic Variables and Equations}

It was shown in the previous subsection
that $\tmur \sim (\tEc1)^2 \sim 10^{-5}$   
when the system is uniform along $x$ 
and the global filling factor is an even integer.  
Since $\tmur \ll 1$ ($\mur \ll \hbar \oc$) and 
$\tmur \ll \tTe$ ($\mur \ll \kB\Te$) , 
we neglect $\tmur$ in the HDEQs
and use eqs.(\ref{eq:nt}), (\ref{eq:Ut}),  (\ref{eq:jn}), and  (\ref{eq:jU}). 
Macroscopic variables $\Te (y)$ and $E_y (y)$ 
are obtained by solving the two HDEQs, 
and $\mur (y)$, on the other hand, is determined 
by eqs.(\ref{eq:sigma}) and (\ref{eq:Ey}) in the previous subsection,   
using the obtained $E_y (y)$. 
In the present case, $E_x$ does not depend on $y$, either, 
since $\partial E_x /\partial y=\partial E_y /\partial x =0$. 
The constant value of $E_x$ is determined by 
the total current through the system.

\subsection{Equations for $\Te (y)$ and $E_y (y)$} 

The equation of the charge conservation in the present case is 
$\partial j_{ny} / \partial y =0$, 
and the boundary condition is $j_{ny}=0$ at sample edges ($y= 0, W$). 
Then we obtain 
%====================================================
\begin{eqnarray}
\cj_{nx} &=& - (\trxx)^{-1} \cE_x   \ , 
\label{eq:jnx} \\
\cj_{ny} &=& - \tsyy \cE_y - \tsyx \cE_x  =0 \ , 
\label{eq:jny}
\end{eqnarray}
%====================================================
for $0<y<W$, 
with $\trxx=\tsxx/(\tsxx^2+\tsxy^2)$. 
The equation of the energy conservation in the present case is  
%====================================================
\begin{eqnarray}
\cnablay \tk_{\rm h}(\tTe) \cnablay \tTe
&=&- \eta(\tTe) \cnablay \tTe  + F(\tTe)  \ , 
\label{eq:T(y)} \\
\eta(\tTe)&=& {\textstyle {1 \over 2\pi}} \cE_x f_1 (1-f_1) \tTe^{-2} \ , \\
F(\tTe) &=& - \trxx^{-1} \cE_x^2  + \Ch \gamma^{-1} \tTe  \ . 
\end{eqnarray}
%====================================================
Here we have used eq.(\ref{eq:jny}) to eliminate $\cE_y$. 
The boundary condition is 
$j_{Uy}=0$ at $y=0,W$, 
and is given by
%====================================================
\begin{equation}
\tk_{\rm h} \cnablay \tTe = - {\textstyle {1 \over \pi}} f_1 \cE_x 
\ .
\label{eq:BC}
\end{equation}
%====================================================

\subsection{Current Density Modulation}

The above boundary condition eq.(\ref{eq:BC}) shows that 
$\Te$ depends on $y$ due to a nonzero $E_x$.  
This modulation of $\Te$ gives a modulation of $j_x=(-e)j_{nx}$, 
since $\trxx$ in eq.(\ref{eq:jnx}) depends on $\Te$ 
(when $\trxx$ is a constant, on the other hand, 
$j_x$ becomes uniform as pointed out by Thouless\cite{Thouless85}). 
The present current density modulation originates 
from the drift component of the energy flux perpendicular to $\vE$ 
(the second term in eq.(\ref{eq:jU})).  
This energy flux has nonzero $y$ component 
when $E_x \not= 0$  
and produces the modulation of $\Te$ in eq.(\ref{eq:BC}), 
and consequently the modulation of $j_x$.  

\begin{figure}
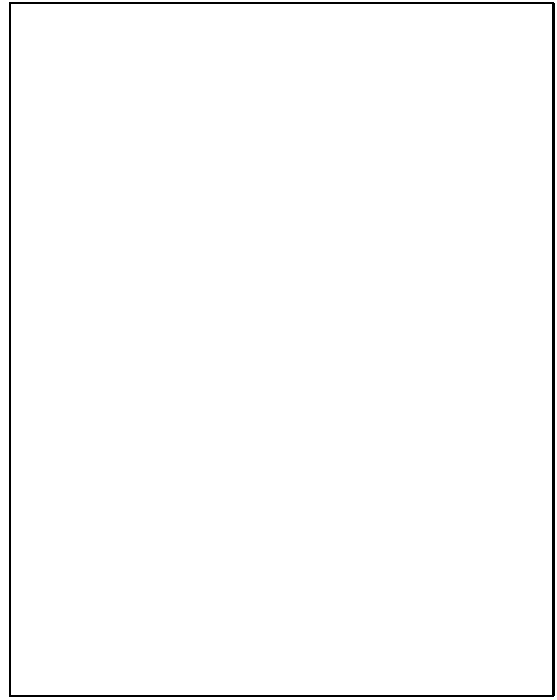

\figureheight{9cm}
\caption{
(a) ``Potential" in a mechanical analogue 
as a function of the ``position" $\tTe$ 
in the vicinity of the higher-$\Te$ stationary point. 
(b) Schematic spatial variations of the electron temperature $\Te$
and the current density $j_x$ in the sample (shaded region) 
when $j_x>0$. 
} 
\label{fig:j_x(y)}
\end{figure}

Since the equation for $\tTe$ is nonlinear, 
numerical calculations are necessary to obtain the solution,  
which will be performed in the future work.  
Here we instead discuss the qualitative profile of $\Te$ and $j_x$ 
by regarding eq.(\ref{eq:T(y)}) as an equation of motion of a particle 
with the ``position" $\tTe$ as a function of the ``time" $\check y$.
In this mechanical analogue, 
$\tk_{\rm h}$ is the ``mass", $\eta$ is the ``friction coefficient", 
and $F$ is the ``force". 
The corresponding ``potential" 
around the higher-$\Te$ stationary point 
is plotted schematically in 
Fig.~\ref{fig:j_x(y)}(a).
The boundary condition eq.(\ref{eq:BC}) gives 
the ``velocity" of the particle at the ``time" $\check y=0,W/\lc1$. 

When $W$ is large, 
the particle spends the most of its ``time" at the top of the ``potential". 
This means that $\Te (y)$ and $j_x (y)$ are approximately uniform 
in the middle of the sample, 
as shown schematically in Fig.~\ref{fig:j_x(y)}(b).
The length scale for the relaxation of $\Te (y)$ and $j_x (y)$ 
from their deviations at the edges is 
of the order of $\lc1\sim 2\mu{\rm m}$.

In the QHE regime with negligible $\rxx$, 
MacDonald {\it et al.}~\cite{MacDonald83} have shown that 
$j_x(y)$ is enhanced near the both edges, 
because of the edge electrostatics in the 2DES. 
When $\rxx$ is large, on the other hand, 
we find in Fig.~\ref{fig:j_x(y)}(b) that 
$j_x(y)$ in one edge is enhanced, 
while $j_x(y)$ in the other edge is suppressed. 
This comes again from the boundary condition eq.(\ref{eq:BC})  
stating that the derivative of $\Te (y)$ has the same sign in the two edges. 
%#############################################################
%#############################################################

\section{Conclusions and Discussions} 

We have derived hydrodynamic equations (HDEQs)
for quantum Hall systems in the regime of large energy dissipation. 
We have shown that 
the strong magnetic field in the present systems produces 
a component of the energy flux 
which is perpendicular to the electric field. 
The presence of this energy flux is the most important feature of 
our HDEQs for the quantum Hall systems 
in the high-electron-temperature regime.  
This energy flux is absent in 
the equation proposed by the previous authors.~\cite{Gurevich84} 
The HDEQs with this unique energy flux 
have predicted a new type of current density modulation. 

Another example of the application of the HDEQs is 
to investigate  
spatial evolutions of $\Te$ in the current direction  (along $x$)
when a spatial transition occurs 
between the small $\rxx$-state and the large $\rxx$-state along $x$. 
The observed spatial evolutions of $\rxx$ in such transitions 
have exhibited a divergence of the relaxation length 
as the current density approaches 
a critical value.~\cite{Kawaguchi97,Kaya98,Kaya99}  
In our previous paper,~\cite{Akera01}
we have applied the previous one-dimensional HDEQ to this problem and 
shown that the calculated relaxation length also exhibits a divergence 
qualitatively similar to the observed one. 

A hydrodynamic model\cite{Eaves01} 
has been proposed to explain 
the observed longitudinal voltage steps.\cite{Cage93,Eaves00}
It is argued in this model that 
electron-hole pairs created at inter-Landau-level scatterings 
near a charged impurity
have common features with 
vortices created when a classical fluid flows past an obstacle. 
The hydrodynamic equations in the present paper, 
on the other hand, describe macroscopic spatio-temporal variations, 
and their typical length scale is much larger than 
the size of a single electron-hole pair.

\section*{Acknowledgements}

The author would like to thank T.~Ando, H.~Aoki, 
Y.~Levinson, N.~Tokuda, and Y.~Asano for valuable discussions. 
This work was supported in part by 
the Grant-in-Aid for Scientific Research (C)
from Japan Society for the Promotion of Science, 
and in part by the Visiting Researcher's Program 
of the Institute for Solid State Physics, the University of Tokyo.

%++++++++++++++++++++++++++++++++++++++++++++++++++++

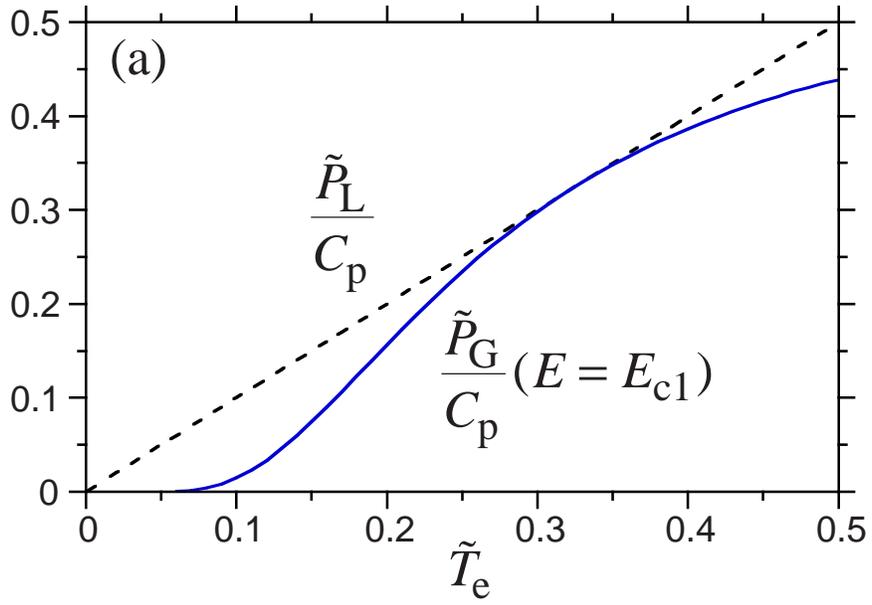
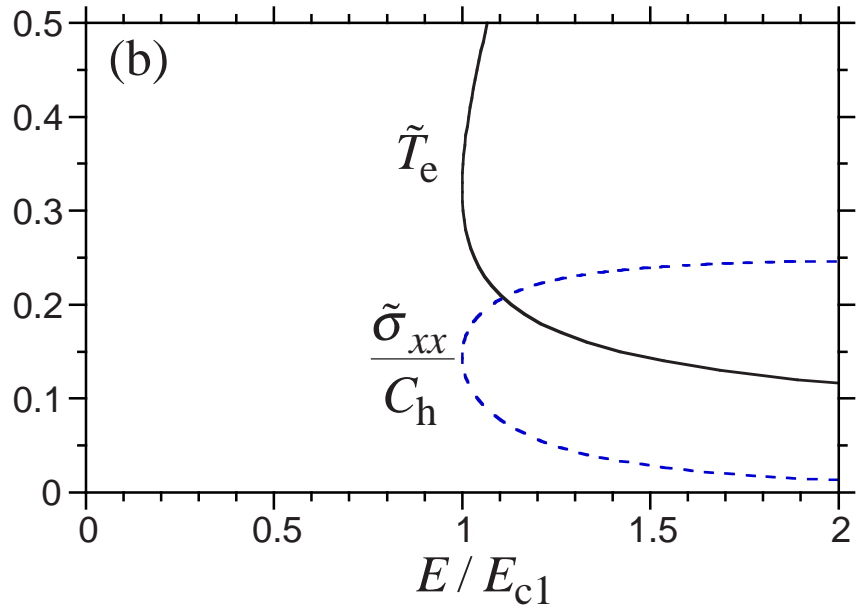
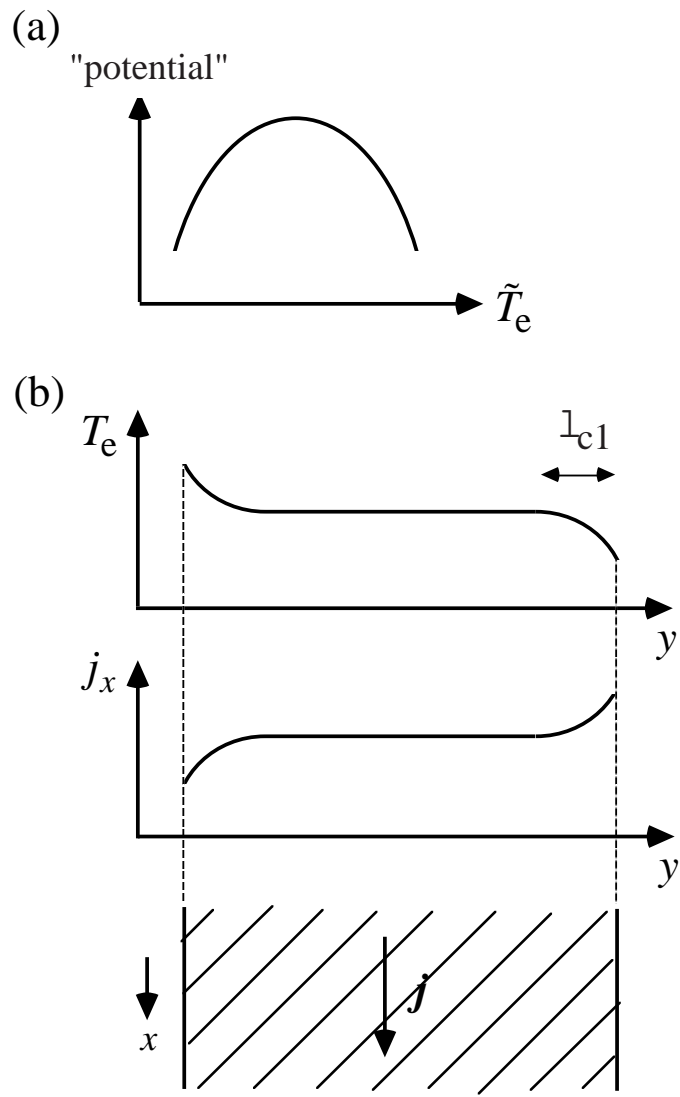
\end{document}